\documentstyle[aps,pra]{revtex}
\def\be{\begin{equation}}
\def\ee{\end{equation}}
\begin{document}
\newcommand\beq{\begin{equation}}
\newcommand\eeq{\end{equation}}
\newcommand\bea{\begin{eqnarray}}
\newcommand\eea{\end{eqnarray}}

\def\n{{\otimes n}}
\def\hcal{{\cal H}}
\def\prawo{\rightarrow}
\def\ra{\rangle}
\def\la{\langle}
\def\eps{\epsilon}
\newcommand{\ereg}{E^\infty_f}
\newcommand{\ket}[1]{| #1 \rangle}
\newcommand{\bra}[1]{\langle #1 |}
\newcommand{\braket}[2]{\langle #1 | #2 \rangle}
\newcommand{\proj}[1]{| #1\rangle\!\langle #1 |}
\newcommand{\ba}{\begin{array}}
\newcommand{\ea}{\end{array}}
\newtheorem{theo}{Theorem}
\newtheorem{defi}{Definition}
\newtheorem{lem}{Lemma}
\newtheorem{exam}{Example}
\newtheorem{prop}{Property}
\newtheorem{propo}{Proposition}
\newtheorem{cor}{Corollary}
\newtheorem{conj}{Conjecture}


\author{Patrick M. Hayden$^1$, Micha\l{} Horodecki$^2$ and Barbara M. Terhal$^3$}

\title{The Asymptotic Entanglement Cost of Preparing a Quantum State}

\address{\vspace*{1.2ex}
        \hspace*{0.5ex}{$^1$Centre for Quantum Computation,
        Clarendon Laboratory, Parks Road, Oxford, OX1 3PU, UK;\,}\\
        \hspace*{0.5ex}{$^2$Institute of Theoretical Physics and Astrophysics,
        University of Gda\'{n}sk, 80-952, Gda\'{n}sk, Poland;\,}\\
        \hspace*{0.5ex}{$^3$IBM Watson Research Center,
        P.O. Box 218, Yorktown Heights, NY 10598, US.\,}\\
        Emails: {\tt
        patrick.hayden@qubit.org,
        michalh@iftia.univ.gda.pl,
        terhal@watson.ibm.com}}

\date{\today}

\maketitle
\begin{abstract}
We give a detailed proof of the conjecture that
the asymptotic entanglement cost of preparing a state
$\rho$ is equal to $\lim_{n \rightarrow \infty}E_f(\rho^{\otimes n})/n$ where
$E_f$ is the entanglement of formation.
\end{abstract}


\section{Introduction}

One of the central issues in quantum entanglement theory is to
determine how to optimally convert between different entangled states
shared by distant observers Alice and Bob.
More precisely, one is interested in the conversion of
$m$ pairs of particles, with each pair in a state $\rho$,
into $n$ pairs, each in another state $\rho'$, by means of
local quantum operations and classical communication (LOCC)
\cite{bdsw}, so that the asymptotic ratio $m/n$ is minimal.
Of course, the perfect transformation by LOCC 
 \[
\rho^{\otimes m} \rightarrow \rho'^{\otimes n}
\]
is usually impossible. Therefore, one permits imperfections and
requires only asymptotically perfect transformations: the state
$\rho^{\otimes m}$ is transformed into some state
$\rho'_n$, which for large $n$ approaches
$\rho'^{\otimes n}$ with a chosen distance measure $D$:
\[
\lim_{n\prawo\infty} D(\rho'_n,\rho'^\n)=0.
\]
If the final state $\rho$ is a two-qubit singlet state $\frac{1}{\sqrt{2}}\left(\ket{01}-\ket{10}\right)$(which we will denote by $\ket{\Psi_-}$),
then the process of conversion is called {\it distillation} \cite{Bennett-dist}.
If, instead, it is the initial state that is in the form of singlets, then one refers to {\it formation} \cite{bdsw}.
In this paper we are interested in the latter process.
We will call the optimal asymptotic yield the entanglement
cost, and denote it by $E_c$.
In Ref. \cite{bbps} it was shown that for a pure state
$\rho=|\psi\ra\la\psi|$, $E_c$ is equal to the entropy of either
of its reductions, e.g. $\rho_A = {\rm Tr}_B(\rho)$.
Thus, to produce $\ket{\psi}^\n$ one needs
$m\approx nS(\rho_A)$  singlets, i.e. the initial state has to be
${\ket{\Psi_-}}^{\otimes n S(\rho_A)}$.

That result suggested the following stochastic method for the
production of $\rho$ out of singlets \cite{bdsw}. One decomposes $\rho$ into
an ensemble $\rho=\sum_ip_i|\psi_i\ra\la\psi_i|$ of pure states.
Then one picks a state
$\ket{\psi_i}$ according to the probability distribution $\{p_i\}$, makes
the state $\ket{\psi_i}$ from initially shared singlets, and finally forgets
the identity of the state.
Therefore, one needs on average $\sum_ip_i S(\rho^A_i)$
singlets (where $\rho_i^A$ is the reduction of $\ket{\psi_i}$). One can choose
the ensemble which minimizes the above average, with the corresponding
minimal cost being called the entanglement of formation of $\rho$ \cite{bdsw},
which we will denote by $E_f(\rho)$.

The above scenario can be improved if we realize that it might be more
economical to produce the state $\rho^\n$ all at once than it is to produce
its $n$ constituents one by one
(see e.g. \cite{woot}).
Thus, the proposed optimal cost should be revised to \cite{acta}
\beq
E_f^\infty(\rho)=\lim_{n\prawo\infty}{E_f(\rho^\n)\over n}.
\eeq
(In the following we will argue that the limit on the right-hand-side of this 
equation exists.)
This quantity is believed to be equal to the entanglement cost
of preparing the state $\rho$, $E_c(\rho)$.
The definition of the entanglement cost, however, refers
to a deterministic number of input singlets, while in the stochastic
method the number of input singlets is a random variable.
Furthermore it is not clear that the stochastic method is really the optimal
way to produce the state $\rho$. In this paper we resolve these issues
by proving that $E_f^\infty(\rho)$ is equal to the entanglement
cost $E_c(\rho)$.
Our result is in a sense dual to that of Rains \cite{Rains} concerning
the entanglement of distillation. He showed that the entanglement of distillation,
if defined in a way analogous to our definition of the entanglement cost,
is equal to the expected yield of a stochastic protocol for generating
output singlets.

To begin, let us sketch our approach.  In our proof, we will first
show that $E_f^\infty(\rho) \geq E_c(\rho)$, i.e. that there exists a formation protocol
that achieves the asymptotic rate ${m\over n}\approx E^\infty_f(\rho)$,
by explicitly constructing such a protocol based
on the law of large numbers in both its classical
and quantum \cite{Schumacher} forms.  Next, we will show that $\ereg(\rho) \leq E_c(\rho)$. The latter inequality will be derived using only some
general properties of the entanglement of formation, such as
its monotonicity under LOCC operations and its quite strong continuity
properties, proved by Nielsen \cite{nielsen:continuity}.
Such an approach, focussing on abstract features of quantum
information-theoretic quantities,
has already proven  to be fruitful in the domain
of quantifying entanglement as well as in the study of quantum channel capacities
\cite{miary,dwie}, providing a simple view of intrincate topics.
To illustrate its power we will now  sketch the proof of the inequality
$\ereg(\rho) \leq E_c(\rho)$.

As mentioned before, the entanglement of formation is monotonically decreasing
under LOCC operations \cite{bdsw}, i.e.
\beq
E_f(\rho)\geq E_f({\cal L}(\rho)),
\eeq
for any state $\rho$ and LOCC operation ${\cal L}$. Moreover, the entanglement
of formation is continuous \cite{nielsen:continuity} in the sense that for
states $\rho$ and $\rho'$  we have
\begin{equation}
\left|E_f(\rho)-E_f(\rho')\right|
\leq 5 D(\rho,
\rho') \log_2 \dim {\cal H} + 2 \eta( D(\rho,\rho') ),
\label{eqnielsencont}
\end{equation}
where $\rho$ and $\rho'$ are supported on the Hilbert space ${\cal H}$
and $\eta(x) = -x \log_2 x$,
under the assumption that $D(\rho,\rho')$
is sufficiently small. $D$ is the Bures distance, given by
$D(\rho,\rho')=2\sqrt{1-F(\rho,\rho')}$ with $F(\rho,\rho')
={\rm Tr}\sqrt{\rho^{1/2} \rho' \rho^{1/2}}$ (Note that the normalization factor for $D$ is not completely standardized.  We have made our choice to agree with Ref. \cite{nielsen:continuity}). The function $F$ is called the Uhlmann fidelity or square-root-fidelity \cite{uhlmann_fid,Jozsa}. Inequality (\ref{eqnielsencont}) implies that if two states
are close to one another, then so are their {\it densities} of entanglement.
A similar continuity result has been proved in
Ref. \cite{Donald} for the relative entropy of entanglement \cite{VP}.

Now, in order to prove $\ereg\leq E_c$, consider the optimal formation protocol,
i.e. the sequence of LOCC operations $\Lambda_n$ producing
states $\rho_n\approx \rho^\n$ out of ${\ket{\Psi_-}}^{\otimes m}$, so that
$m/n$ tends to $E_c$. If we take large $n$, so that $D(\rho^\n,\rho_n)$
is small, then by continuity (\ref{eqnielsencont}) we have
\beq
{E_f(\rho^\n)\over n}\approx {E_f(\rho_n)\over n}
\label{eq1}
\eeq
Since $\rho_n=\Lambda_n({\ket{\Psi_-}}^{\otimes m})$, then from the
monotonicity of $E_f$ we obtain
\beq
{E_f(\rho_n)\over n} \leq {E_f({\ket{\Psi_-}\bra{\Psi_-}}^{\otimes m})\over n}={m\over n}.
\label{eq2}
\eeq
Now, by defnition, the left-hand-side of the equation (\ref{eq1}) tends
to $\ereg$ while the last term of estimate (\ref{eq2}) tends to $E_c$,
hence we obtained the required inequality.

Since we used only two properties of the entanglement of formation, we can
rephrase the result in a more general setting. Consider any function $f$, which
can be regularized, i.e. for which the limit
$f^\infty(\rho)=\lim_{n \rightarrow \infty} {f(\rho^\n)\over n}$ exists. Now, if $f$ is monotonic
and continuous in the sense of equation (\ref{eqnielsencont}), then
$f^\infty$ is a lower bound for $E_c$.
This supports the view, according to
which the dual measures of entanglement, entanglement of distillation $E_D$
and entanglement cost $E_c$ are in a sense extreme ones \cite{miary}.

\section{Entanglement cost and entanglement of formation}

Let us now pass to the fully rigorous part of the paper.
Analogous to the definition of asymptotic reducibility in
Ref. \cite{multibenn} (cf. \cite{Werner}), we define the asymptotic
entanglement cost
for the preparation of a bipartite state to be
\beq
E_{c}(\rho)=\inf \left\{E\,|\,\forall \eps>0, \delta > 0,\, \exists\, m,n,{\cal L}, \right.
\,|E-\frac{m}{n}| \leq \delta  \mbox{        and         }
\left.
D({\cal L}(\ket{\Psi_-}\bra{\Psi_-}^{\otimes m}),\rho^{\otimes n}) \leq \eps \right\},
\label{defent}
\eeq
where $\ket{\Psi_-}$ is the singlet state in ${\bf C}_2 \otimes
{\bf C}_2$, ${\cal L}$ is an LOCC superoperator and $D$ is again
the Bures distance.

Our main result is the following:

\begin{theo}
The asymptotic entanglement cost of preparing a state $\rho$ is given by
\beq
E_{c}(\rho) = \lim_{n \rightarrow \infty} \frac{E_f(\rho^{\otimes n})}{n},
\eeq
where $E_{c}(\rho)$ is defined in Eq. (\ref{defent})
and $E_f(\rho)$ is the entanglement of formation of $\rho$,
defined as
\beq
E_f(\rho)=\min_{{\cal E}=\{p_i,\ket{\psi_i}\}} \sum_i p_i E(\ket{\psi_i}\bra{\psi_i})
\label{defeof},
\eeq
and $E(\ket{\psi}\bra{\psi})$ is the von Neumann entropy of the reduced
density matrix of $\ket{\psi}$.
\label{theo_asym}
\end{theo}

To make sense of the above claim, we note that the sequence
$\left(\frac{E_f(\rho^{\otimes n})}{n}\right)$ has a well-defined
limit.  This is a consequence of the
fact that if a sequence $(a_n)$ satisfies
$a_n\leq cn$ for some constant $c$ and $a_n+a_m\geq a_{n+m}$ for any $m,n$,
then $(a_n/n)$ is convergent \cite{BNS}. It is easy to see that our sequence
satisfies the conditions.

We begin by proving that $E_c$ obeys a form of additivity that
is highly desirable for an asymptotic cost function,
and that will be useful in what follows.

\begin{lem}
$E_c(\rho^{\otimes k}) = k E_c(\rho)$ for all $k=1,2,\ldots$.
\label{lemadditive}
\end{lem}
{\em Proof}
One direction is simple: since a protocol to approximate $\rho^{\otimes k}$
is just a protocol to approximate $k$ copies of $\rho$,
$E_c(\rho^{\otimes k}) \geq k E_c(\rho)$.  The first step in demonstrating
that the inequality holds in the opposite direction will be to show that
in the definition of $E_c$, $n$ can be taken to be arbitrarily large.
Suppose not, in other words, that for fixed $\epsilon, \delta > 0$,
it is impossible to choose $n \geq N$, $m$, and ${\cal L}$ such that
$| E_c(\rho) - m/n | \leq \delta$ and
$D({\cal L}(\proj{\Psi_-}^{\otimes m}),\rho^{\otimes n}) \leq \epsilon$.
By the definition of $E_c(\rho)$ and the fact that the set of LOCC
operations is closed, this implies, however, that there exist
$n < N$, $m$, and ${\cal L}$ such that
$E_c(\rho)=m/n$ and ${\cal L}(\proj{\Psi_-}^{\otimes m}) = \rho^{\otimes n}$.
Setting $r = \lceil N/n \rceil$ and applying ${\cal L}^{\otimes r}$
to $\proj{\Psi_-}^{\otimes mr}$ then violates the assumption that
$n$ could not be taken arbitrarily large.


Now, to complete the proof, we will pick $n \gg k$ in order to efficiently
approximate some number of copies of $\rho^{\otimes k}$.  Formally,
we write $n=rk+s$,
where $r$ and $s$ are non-negative integers and $s<k$.  If we produce
an approximation to within $\epsilon$ of $\rho^{\otimes n}$ starting
from $m$ singlets and then
throw away the extra $s$ copies of $\rho$, we are left with an approximation to
within $\epsilon$ of $\rho^{\otimes rk}$ that still required $m$ singlets.
The waste, however, becomes insignicant since
$| E - m/n | \prawo | E - m/rk |$ as $n \prawo \infty$.
$\Box$

Note that full additivity of $E_c(\rho)$, i.e. $E_c(\rho \otimes \sigma)=
E_c(\rho)+E_c(\sigma)$ would be a stronger statement, possibly requiring additivity
of $E_f$. With that technical lemma concluded,
Theorem \ref{theo_asym} becomes a consequence of the following two lemmas.

\begin{lem}
Let the entanglement of formation of a (finite dimensional) density matrix $\rho$ be $E_f(\rho)=
\sum_i p_i E_f(\ket{\psi_i}\bra{\psi_i})$ where the optimal ensemble is given by $\{p_i,\ket{\psi_i}\}_{i=1}^k$. We have $E_{f}(\rho) \geq E_c(\rho)$.
\label{lemecleq}
\end{lem}

{\em Proof} In the limit
of large $n$, we will approximate $\rho^{\otimes n}$ by
\beq
\rho_{T_{\delta_1}^{(n)}}=\sum_{s \in T_{\delta_1}^{(n)}} \ket{\psi_s}\bra{\psi_s},
\eeq
where $T_{\delta_1}^{(n)}$ is the (strongly) typical set defined as follows. All states $\ket{\psi_s}$ (unnormalized) in this set are tensor products of states $\sqrt{p_i} \ket{\psi_i}$ and they are such that every state $\sqrt{p_i}
 \ket{\psi_i}$ occurs ${p_i n \pm \delta_1 n \log_{p_i}(2)/k }$ times in the tensor product.  (We assume that $p_i < 1$ for all $i$.  Otherwise, $\rho$ is a pure state.) This implies that the total probability $p_s=\langle \psi_s | \psi_s \rangle$ for a state $\ket{\psi_s}$ is bounded as
\beq
2^{-n(H(\vec{p}))+\delta_1)} \leq p_s \leq 2^{-n(H(\vec{p}))-\delta_1)},
\eeq
where $H(\vec{p})$ is the Shannon entropy of $\vec{p}=(p_i)$. The density matrix $\rho_{T_{\delta_1}^{(n)}}$ is basically the state $\rho^{\otimes n}$ constructed by cutting off the unlikely sequences of states and then renormalizing \cite{noquantum}. In order to bound the fidelity for this approximation, we add these unlikely sequences of states with probability 0 in the sum. 
For any $\delta_1 > 0$, $\eps_1 > 0$ and $n$ sufficiently large we have by
the asymptotic equipartition theorem \cite{cover} that the total probability $p_{T_{\delta_1}^{(n)}}$ for the typical set is larger than $1-\eps_1$, where $\eps_1 \rightarrow 0$ as $n \rightarrow \infty$. Thus, using
the joint concavity of $F$ \cite{uhlmann_fid}, we can bound
\beq
F(\rho^{\otimes n},\rho_{T_{\delta_1}^{(n)}}) \geq \sqrt{1-\eps_1}.
\label{fid1}
\eeq

Consider the density matrix $\rho_{T_{\delta_1}^{(n)}}$ and its decomposition in terms
of the states $\ket{\psi_s}$. Each state $\ket{\psi_s}$ can be obtained
from a set of EPR pairs by entanglement dilution \cite{bbps}.
In particular, let $\ket{\psi_s}$ be a state
in which every state $\ket{\psi_i}$ occurs $p_i n \pm \delta_1 n \log_{p_i}(2)/k $ times.
Starting from a set of maximally entangled EPR pairs, we do entanglement dilution for
each state $\ket{\psi_i}^{\otimes p_i n \pm \delta_1 n \log_{p_i}(2)/k}$. For any $\delta_2$ and
$\epsilon_2$ greater than zero there exists an $n$ such that starting from
$(p_i n+\delta_1 n \log_{p_i}(2)/k)[E_f(\ket{\psi_i}\bra{\psi_i})+\delta_2]$ EPR pairs we can obtain an approximation to $\ket{\psi_i}^{\otimes p_i n\pm\delta_1 n \log_{p_i}(2)/k}$ which has square-root-fidelity larger than $1-\eps_2$. Since there are
$k$ states in the optimal ensemble (and $k$ is finite), we can therefore approximate the state
$\ket{\psi_s}$ with square-root-fidelity $|\langle \psi|\psi'\rangle| > (1-\eps_2)^k$,
starting from $n \sum_i p_i E_f(\ket{\psi_i}\bra{\psi_i})+n (O(\delta_1)+O(\delta_2))$ EPR pairs, with $\eps_2 \rightarrow 0$, $\delta_1 \rightarrow 0$, and $\delta_2 \rightarrow 0$ for $n \rightarrow \infty$.

The approximation of $\rho_{T_{\delta_1}^{(n)}}$ by $\rho'_{T_{\delta_1}^{(n)}}=\sum_{s \in T_{\delta_1}^{(n)}} \ket{\psi_s'}
\bra{\psi_s'}$, where $\ket{\psi_s'}$ is the approximation of $\ket{\psi_s}$ which we obtain by entanglement dilution starting from the set of EPR pairs, has the property that
\beq
F(\rho_{T_{\delta_1}^{(n)}},\rho'_{T_{\delta_1}^{(n)}}) \geq \sum_{s \in T_{\delta_1}^{(n)}} |\langle \psi_s|\psi_s' \rangle| \geq
(1-\eps_2)^k\equiv 1-\eps_3,
\label{fid2}
\eeq
where $\eps_3 \rightarrow 0$ for $n \rightarrow \infty$, since $k$ is finite.
Furthermore, since we can make every state $\ket{\psi_s'}$ starting from a
given set of EPR pairs, we can make any convex combination, for example
$\rho_{T_{\delta_1}^{(n)}}'$, of the states $\ket{\psi_s'}$ (see \cite{jonplen} and also
\cite{hay_ter_uhl}), starting from this same set of EPR pairs.

Finally, we can use the triangle inequality for the Bures metric, and Eqs. (\ref{fid1}) and
(\ref{fid2}) to obtain that
\beq
D(\rho^{\otimes n},\rho_{T_{\delta_1}^{(n)}}') 
\leq 
2 \sqrt{1-\sqrt{1-\eps_1}}+ 2 \sqrt{\eps_3}.
\label{eqtriangle}
\eeq
which is the desired result. $\Box$

The lemma can be applied to bound $E_c(\rho)$ from above
by $E_f(\rho^{\otimes k})/k$ where $k$ is any fixed number of copies of $\rho$,
using Lemma \ref{lemadditive}.  Consequently, we have that
\beq
E_f^{\infty}(\rho) \geq E_c(\rho).
\eeq

Let us now prove the converse of this relation.

\begin{lem}
$E_f^{\infty}(\rho) \leq E_c(\rho)$.
\label{lemecgeq}
\end{lem}

{\em Proof} The basic idea of the proof, as sketched in the introduction,
is to use the continuity of the
entanglement of formation \cite{nielsen:continuity} and its monotonicity
under LOCC operations. Throughout the proof we will
use the notation $A_n=\frac{E_f(\rho^{\otimes n})}{n}$.
Now suppose that the
lemma does not hold, so that $\lim A_n > E_c(\rho)$.
It follows that there exists an integer $N$, such that for all $k > N$,
\begin{equation}
\left|A_k - \lim A_n\right| < \Delta
= \frac{\lim A_n - E_c(\rho)}{4} > 0.
\label{defDelta}
\end{equation}
Let $\rho_k= {\cal L}(\proj{\Psi_-}^{\otimes m})$ be an
approximation to $\rho^{\otimes k}$.
Now, supposing that $\rho$ is supported on Hilbert space ${\cal H}$,
fix $\epsilon$ such that
$5 \eps \log \dim {\cal H} - 2\epsilon\log\eps=\Delta$.
It then follows from Eqs. (\ref{eqnielsencont}) and our choice of $\epsilon$
that if $D(\rho^{\otimes k}, \rho_k) < \epsilon$ then
\beq
|E_f(\rho^{\otimes k}) - E_f(\rho_k)| < k \Delta.
\label{diffk}
\eeq
Now we will apply the definition of $E_c(\rho)$ to fix $k$. From the definition of $E_c$, Eq. (\ref{defent}), and the proof of
Lemma \ref{lemadditive}, we have that there exists a $k > N$ as well
as $m$ and ${\cal L}$  such that
\beq
|E_c(\rho) - {m \over k}| \leq \Delta
\label{del}
\eeq
and $D(\rho^{\otimes k},\rho_{k}) < \eps$, where
$\rho_k = {\cal L}(\proj{\Psi_-}^{\otimes m})$.

Next, we can bound
\beq
\left| \lim A_n - \frac{E_f(\rho_k)}{k} \right|
\leq  \left| \lim A_n - A_k \right|
        + \left| A_k - \frac{E_f(\rho_k)}{k} \right|
< 2 \Delta,
\eeq
by using Eqs. (\ref{defDelta}) and (\ref{diffk}).

This gives $E_f(\rho_k) > k( \lim A_n - 2 \Delta )$.
On the other hand, by Eq. (\ref{del}), we have
\begin{equation}
E_f(\proj{\Psi_-}^{\otimes m})
= m < k( E_c(\rho) + \Delta ),
\end{equation}
which finally yields
\beq
E_f(\rho_k) - E_f(\proj{\Psi_-}^{\otimes m})
> k( \lim A_n - E_c(\rho) - 3 \Delta )
= k \Delta,
\eeq
a contradiction since the entanglement of formation
cannot increase under LOCC operations.
$\Box$

\section{Alternative definitions of the asymptotic cost}

While $E_c(\rho)$ is perhaps the most natural function to
associate with the asymptotic cost of preparing a bipartite state, other
definitions would have been consistent with our discussion in the
introduction.  An example of a different but perhaps useful definition
of the asymptotic entanglement cost is the following:
\beq
E_{alt}(\rho)=\inf \left\{E\,|\,\forall \eps>0, \delta > 0,\,
\exists  N \, |\, (\forall n > N\, \exists\, m, {\cal L}, \right.
\,|E-\frac{m}{n}| \leq \delta  \mbox{        and         }
\left.
D({\cal L}(\ket{\Psi_-}\bra{\Psi_-}^{\otimes m}),\rho^{\otimes n}) \leq \eps ) \right\}.
\label{defentc}
\eeq

The difference between $E_c$ and $E_{alt}$ is that for fixed fidelity
and cost the former
only requires the existence of a single $n$ such that $\rho^{\otimes n}$
can be approximated efficiently while the second requires that there
exist a threshold $N$ such that for all $n > N$ the state $\rho^{\otimes n}$
can be approximated efficiently.  One consequence of this difference is
that it is immediately clear that $E_{alt}$ is additive in this sense
of Lemma \ref{lemadditive}, while some work was required to prove that
$E_c$ was.  Indeed, a priori, one might not expect the
two definitions to agree.  Suppose, for example, that for fixed $\epsilon$
there exists a protocol to make an approximation $\rho_n$ of
$\rho^{\otimes n}$ from $m$ singlets, with the approximation satisfying
$D(\rho_n,\rho^{\otimes n}) = \epsilon$.  In order to produce an approximation
to $\rho^{\otimes nk}$ from $km$ singlets, one might think that
applying this protocol $k$ times
would be a good strategy.  One finds, however, that for $\epsilon > 0$,
\beq
\lim_{k\prawo\infty}D(\rho^{\otimes nk},\rho_n^{\otimes k})
= \lim_{k\prawo\infty}2\sqrt{1-(1-\epsilon^2)^k}
= 2.
\eeq
The protocol, therefore, generally fails to produce a good approximation for
large $k$.  This example suggests that the existence of a threshold $N$
beyond which approximations to some given fidelity are always possible is a
very difficult condition to satisfy.
Nonetheless, by applying the results of the previous
section, it is actually easy to see that the definitions $E_c$ and
$E_{alt}$ are equivalent, so that the extra condition can be met
without increasing the asymptotic unit cost.
First, notice that the argument of Lemma \ref{lemecleq}
actually also works for $E_{alt}$ so that
$E_{alt}(\rho) \leq E_f^\infty(\rho)$.  Next, since the definition
of $E_{alt}$ is more stringent than that of $E_c$, we have that
$E_c(\rho) \leq E_{alt}(\rho)$.  Combining these inequalities with
the result of Lemma \ref{lemecgeq}, we get
\beq
E_f^\infty(\rho) \leq E_c(\rho) \leq E_{alt}(\rho) \leq E_f^\infty(\rho),
\eeq
so that these two definitions of the entanglement cost always
agree.

A final pair of alternative definitions for the asymptotic entanglement
cost would use the trace distance
\beq
d(\rho,\sigma) = {1 \over 2}{\rm Tr}|\rho-\sigma|,
\eeq
in place of the Bures distance.  Lemma \ref{lemadditive}
is easily seen to hold for $d$ since its proof uses only axiomatic properties
of all metrics and stability with respect to tensor products but no other properties 
specific to $D$.
The status of Lemmas \ref{lemecleq} and \ref{lemecgeq} is slightly more 
involved. The trace distance and
Bures distance are related by the following inequalities \cite{distinguish}
\beq
{1 \over 4} D(\rho,\sigma)^2 = 1 - F(\rho,\sigma) \leq d(\rho,\sigma)
\leq \sqrt{1-F(\rho,\sigma)^2} = \sqrt{{D^2 \over 2} - {D^4 \over 16}},
\label{eqmetricrelate}
\eeq
which show that $d$ and $D$ are equivalent metrics and bounds
them in terms of each other by functions that are independent of the
dimension of the underlying Hilbert space, ${\cal H}$.
In other words, $d$ and $D$ are equivalent metrics even in the asymptotic
regime.
To be concrete, applying the
right-hand inequality to Eq. (\ref{eqtriangle}), for example, is sufficient
to show that Lemma \ref{lemecleq} also holds for the trace distance.
In order to prove Lemma \ref{lemecgeq}, we used the continuity relation,
Eq. (\ref{eqnielsencont}), but effectively only required the weaker
inequality
\beq
| E(\rho) - E(\sigma) | \leq B\,D(\rho,\sigma)\,\log\dim {\cal H} + C,
\eeq
where $B$ and $C$ are constants and ${\cal H}$ is the supporting
Hilbert space.  Applying Eq. (\ref{eqmetricrelate}), however, gives
an inequality of the form
\beq
| E(\rho) - E(\sigma) | \leq B\,\sqrt{d(\rho,\sigma)}\,\log\dim {\cal H} + C,
\eeq
which again is sufficient to carry through the rest of the proof.
Therefore, all our conclusions hold even if the
entanglement cost is defined using the trace distance.  Indeed, they should
hold for any metric equivalent to the Bures metric in which the equivalence
is given by a function that is independent of the dimension of ${\cal H}$.

\section{Conclusions}
We have given two rigorous definitions of the asymptotic cost of
preparing a bipartite mixed state $\rho$ and shown them both to be equal
to the regularized entanglement of formation, $E_f^\infty(\rho)$,
resolving an important conjecture in the theory of quantum entanglement.
Furthermore, we have shown that this asymptotic cost is fairly insensitive
to the choice of metric on density operators.  In particular, the Bures
distance and trace distance result in identical asymptotic costs.

An important problem left open by this work is the question of actually
evaluating $E_f^\infty(\rho)$.  Even the non-regularized function $E_f(\rho)$
is notoriously difficult to calculate; its value is only known for some
very special cases \cite{woot,isotropic}.  If it turns out that $E_f$ is not
additive for tensor products, then, in spite of the results of this paper,
determining the asymptotic cost of preparing a state remains quite a formidable problem.

\section{Acknowledgments} We would like to thank Pawe\l{}
Horodecki, Michael Nielsen and John Smolin for helpful discussions. PMH is grateful to the
Rhodes Trust and the EU project QAIP, contract No. IST-1999-11234, for support. MH acknowledges
support of Polish Committee for Scientific Research, contract No.
2 P03B 103 16 and EU project EQUIP, contract No. IST-1999-11053.
BMT acknowledges support of the ARO under contract
number DAAG-55-98-C-0041.


\begin{thebibliography}{10}
\bibitem{noquantum}Note that we are using an approximation to 
$\rho^{\otimes n}$ which is based on the optimal decomposition for the entanglement 
of formation, and not on the eigendecomposition of $\rho$ 
which is used in Schumacher compression \cite{Schumacher}. Both approximations converge to $\rho^{\otimes n}$ in the limit of large $n$.
\end{thebibliography}
\end{document}